\documentclass[%
 reprint,
 amsmath,amssymb,
 aps,
 pra,
]{revtex4-2}

\usepackage{graphicx}
\usepackage{longtable}
\usepackage{multirow}
\usepackage{longtable}
\usepackage{dcolumn}
\usepackage{bm}
\usepackage{hyperref}
\usepackage{xcolor} %

\usepackage{mathtools}
\usepackage{siunitx}
\usepackage{lineno}
\usepackage{natbib}

\def\njust{Institute of Ultrafast Optical Physics, Department of Applied Physics, Nanjing University of Science and Technology, Nanjing 210094, China}

\def \jkmol {\si{\joule\per\kelvin\per\mole}}
\def\fcc{{\textit{fcc}}}
\def\vib{\mathrm{vib}}
\def\td{{\Theta_{\mathrm{D}}}}

\begin{document}


\title{First-principles thermal equation of state of \fcc{}  iridium}

\author{Kai Luo}
\email{kluo@njust.edu.cn}
\affiliation{\njust}%

\author{R. E. Cohen}%
\affiliation{Extreme Materials Initiative, Earth and Planets Laboratory, Carnegie Institution for Science,
5241 Broad Branch Rd., N.W., Washington, DC 20015, USA}%


\author{Ruifeng Lu}
\affiliation{\njust}%


\date{Modified on Jan 5, 2023} 

\begin{abstract}
The thermal equation of states for \fcc{} iridium (Ir) is obtained from
first-principles molecular dynamics up to 3000 K and 540 GPa. The equation of state (EoS) 
is globally fitted to a simplified free energy model and various parameters 
are derived. The theoretical principal Hugoniot is compared with shockwave experiments,
where discrepancy suggests formation of new  Ir phases. A few representative EoS parameters,
such as bulk modulus $K_T$, thermal expansivity $\alpha$, Gr\"uneisen parameter $\gamma$, and 
constant pressure capacity $C_P$, Debye temperature, $\td$ are computed to 
compare with experimental data. 
\end{abstract}

\maketitle


\section{Introduction}
Iridium (Ir) is a 5d transition metal of the platinum group.
It is the second-densest metal with a density of 22.56 g/cm$^3$ at ambient condition,
only slightly lower by about 0.12\% than the densest metal osmium (Os). It has the largest shear modulus,
 G=210 GPa among the face-centered cubic (\fcc{}) metals.
The solid Ir remains in the \fcc{} structure up to the melting point of 2719 K\cite{Arblaster2010}.
Due to its prominent thermophysical and mechanical properties and high corrosion resistance, it is used in many technological
applications, such as crucibles, thermocouples,  spark plugs, aircraft engine parts, and deep water pipes. 
The lack of phase transitions, simple \fcc{} structure, high melting temperature, and non-reactivity, 
makes it ideal for experiments as a heater, absorber, or standard  for example in diamond-anvil cell (DAC) experiments, 
and ideal for studying effects of compression on noble metals. Our understanding of the properties of Ir is still limited, and fundamental 
research on it remains of great interest. 

With the advances in lab technologies, extreme conditions ($P > 200$ GPa, $T > 2000$ K) become more and more amenable to study. 
Fundamental to all studies at extreme conditions is the equation of state (EoS) that relates $P, V, T$, and $U$ or $F$,
where symbols of $P, V, T, U$, and $F$ stand for pressure, volume, temperature, internal energy, and Helmholtz free energy.
The earliest investigation of iridium EoS dates back to 1937 by P.W. Bridgman 
up to 7 GPa\cite{Bridgman1937,Bridgman1952}, followed by work of Schock and Johnson \cite{Schock1971} and 
then of Akella\cite{Akella1982} up to 30 GPa. 
Cerenius and Dubrovinsky \cite{Cerenius2000} measured the compressibility of Ir using DAC up to 65 GPa. 
Later, Cynn et al. found that Ir has the second-lowest compressibility of any element after Os from their DAC experiment 
up to 65 GPa, which was corroborated by first-principles calculations\cite{Cynn2002}.

For the EoS diagrams, zero-temperature first-principles EoS can be supplemented with 
finite-temperature vibrational
entropies from the phonon dispersions. Phonon frequencies can be calculated from 
finite differences, or with the density-functional perturbation theory (DFPT) \cite{Baroni2001}.
Thanks to the development in the density functional theory toolkit, theoretical EoS for Ir appeared 
in several experimental work \cite{Cynn2002,Burakovsky2016,Monteseguro2019,Fang2010,Kaptay2015}.
However, these theoretical EoS's were limited to low temperatures (around 300 K) using static calculations 
fitted to Birch-Murnaghan (BM) EoS \cite{Birch1947}. Anharmonic lattice vibrations were considered
in Ref.\cite{Burakovsky2016}, but the focus was the phase diagram and phase stability. Anzellini et al.
studied Ir up to 80 GPa and 3100K combining in situ synchrotron X-ray diffraction using laser-heating DACs and
density functional theory calculations \cite{Anzellini2021}.  A comprehensive study covering a larger 
range of temperatures and pressures has not been performed. Indeed, studying other phases would be interesting, but in applications as a standard
in experiments, we focus on the \fcc{} phase.
In this work, we aim to provide the EoS for \fcc{} Ir up to 3000 K and 540 GPa in first-principles molecular dynamics (FPMD).


\section{Theoretical EoS from FPMD}

\label{sec:theoreticalEOS}
First principles methods have been widely adopted in the simulation of condensed phases where no phenomenological parameters
are needed. They give access to a space of thermodynamic conditions, which are hard to reach for experimental 
efforts and can be used to help calibrate experiments, where, for example temperature data are not sometimes available at the desired conditions.
FPMD takes into account of anharmonic 
vibrations of ions directly at finite temperatures through thermostatting. The electronic free energy is given by the 
Mermin-Kohn-Sham density functional theory (DFT) \cite{KohnSham1965,Mermin1965}. FPMD becomes the most used tool for predicting the thermal EoS, 
subject to the exchange-correlation free energy functional approximations\cite{Karasiev2014,Karasiev2016,Karasiev2018}.
Classical molecular dynamics is suitable for high temperatures above the Debye temperature 
as it includes anharmonicity exactly, unlike other approaches. 
At lower temperature deviations from the $P-V-T$ equation of state are small, but heat capacities 
and high order properties such as thermal expansivity which show strong quantum effects at temperature 
below the Debye temperature are indeed less accurate.

\subsection{FPMD details}
\begin{figure}
	\includegraphics[width=1.0\linewidth]{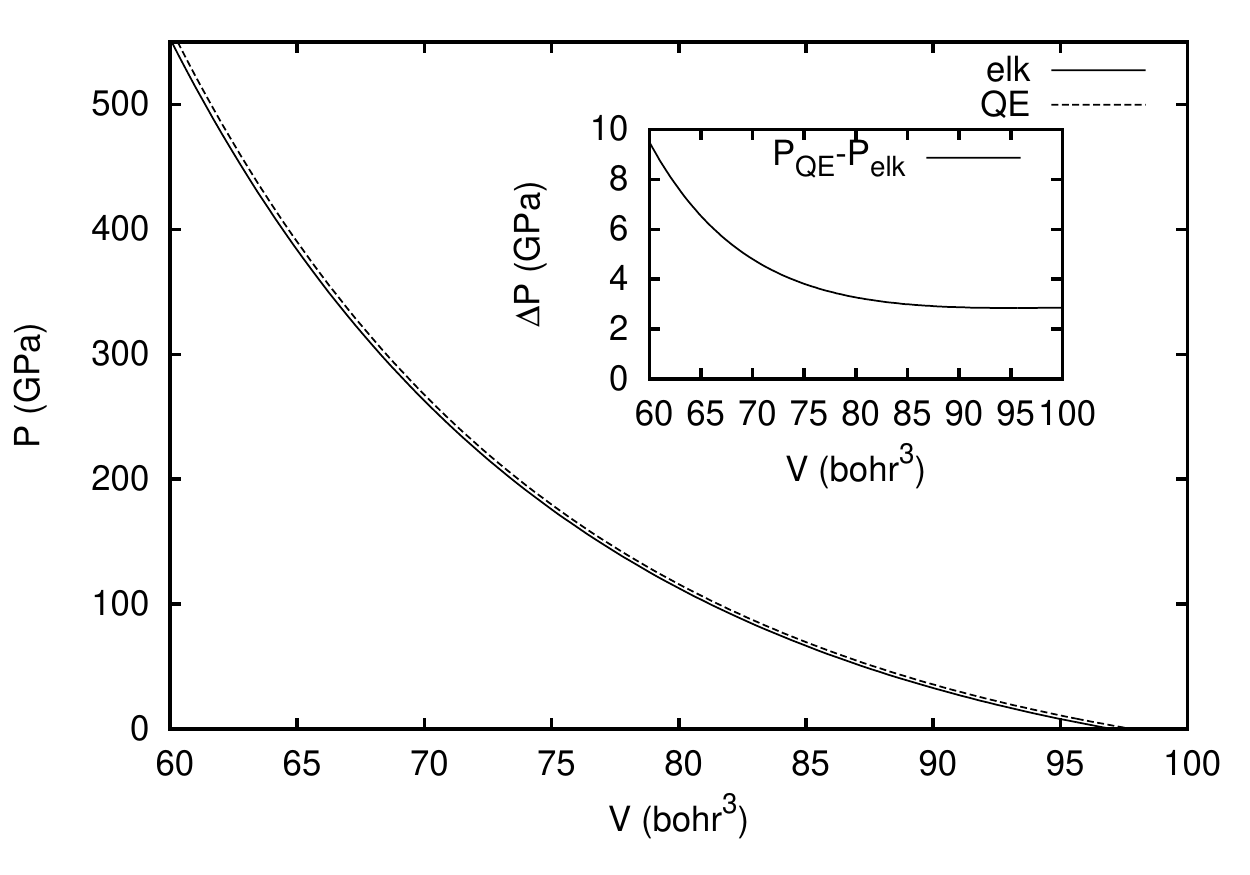}
    \caption{ \label{fig:pseudolapw} Static equations of state from GBRV pseudopotential planewaves (QE) and LAPW (elk) calculation are compared. The Vinet EoS was
    used to fit the energy-volume curve. Inset figure shows the pressure difference and the maximum is less than 10 GPa.
    }
\end{figure}

First, we computed the static EoS of Ir at zero temperature. We used \textit{Quantum Espresso} ver. 6.7 throughout this work \cite{Giannozzi2009}. 
We used the scalar-relativisitic (Garrity-Bennett-Rabe-Vanderbilt) GBRV ultrasoft pseudopotential \cite{GBRV2014} 
with Perdew-Burke-Erzernhorf (PBE) exchange-correlation (xc) functional \cite{PBE1996,PBE1996Erratum}. The electronic configuration for the pseudopotential is $[Xe]5p^{6.0}5d^{8.5}$. 
EoS was derived by fitting energy-volume curve in the 3rd-order BM equation. To validate the range of applicability of the pseudopotential, we performed similar
calculations in linearized augmented planewave (LAPW) code Elk \cite{elk} and above two $P-V$ curves agree well up to 550 GPa (see Fig.\ref{fig:pseudolapw}). 

For the FPMD calculation, we prepared a cubic box containing 108 atoms in the \fcc{} structure. 
The energy cutoffs for planewaves and density are 80 Ry and 320 Ry, respectively. Energy is converged within 5 meV per atom. Only $\Gamma$ point was sampled. 
The bands are occupied according to the Fermi-Dirac distribution at each temperature, and the number of bands are large enough to guarantee the occupation number is smaller than $10^{-7}$ for the highest occupied state.
Early studies showed that the spin-orbit coupling does not affect the EoS and hence we used spin-unpolarized DFT neglecting spin-orbit coupling. 
Then conditions at a combination of lattice constants
$a/a_0= 0.86, 0.88, 0.90, 0.92, 0.96, 1.00$ ($a_0=3.801$ \AA) 
and temperatures $T=300, 1000, 1500, 2000, 2500, 3000$ K were used in the simulations (see conditions in Table \ref{tab:EOS_data}).
The time step is 20 a.u. (0.9676 fs). The equilibrated time steps are more than 2000 to get the statistical means and standard deviations, which
give less than $1\%$ standard deviation. 
The ionic temperature is regulated by the stochastic-velocity rescaling thermostat \cite{Bussi2007} and no quantum corrections to the ionic motion are included.

\begin{table*}[!htp]
    \caption{Pressure $P$ and internal energy per atom $U$ and its standard deviation of the means $P_{\mathrm{err}}$ and $U_{\mathrm{err}}$
    are extracted from FPMD simulations of \fcc{} Ir for a given temperature $T$ and atomic volume $V$ (or the mass density $\rho$). The global minimum of $U$ is 
    underlined.
    }
    \begin{ruledtabular}
        \label{tab:EOS_data}
        \begin{tabular}{rrrrrrr}
             $T$ (K) &  $V$ (bohr$^3$) &  $\rho$ (g/cm$^3$) &  $P$ (GPa) &  $P_{\mathrm{err}}$ (GPa) &  $U$ (Ry) &  $U_{\mathrm{err}}$ (Ry) \\
             \hline
             300 &           61.01 &              35.31 &      527.6 &             0.01 &  -181.353 &         0.00004 \\
             300 &           63.14 &              34.12 &      451.6 &             0.01 &  -181.423 &         0.00004 \\
             300 &           67.54 &              31.89 &      326.1 &             0.02 &  -181.538 &         0.00005 \\
             300 &           72.14 &              29.86 &      229.5 &             0.01 &  -181.624 &         0.00004 \\
             300 &           81.97 &              26.28 &       99.3 &             0.02 &  -181.729 &         0.00005 \\
             300 &           92.65 &              23.25 &       25.0 &             0.03 &  \underline{-181.770} &         0.00006 \\
             \\
            1000 &           61.01 &              35.31 &      531.3 &             0.04 &  -181.339 &         0.00012 \\
            1000 &           63.14 &              34.12 &      455.3 &             0.06 &  -181.410 &         0.00021 \\
            1000 &           67.54 &              31.89 &      330.0 &             0.04 &  -181.524 &         0.00014 \\
            1000 &           72.14 &              29.86 &      233.4 &             0.06 &  -181.611 &         0.00017 \\
            1000 &           81.97 &              26.28 &      103.5 &             0.06 &  -181.715 &         0.00016 \\
            1000 &           92.65 &              23.25 &       29.5 &             0.07 &  -181.756 &         0.00018 \\
            \\
            1500 &           61.01 &              35.31 &      533.9 &             0.08 &  -181.329 &         0.00039 \\
            1500 &           63.14 &              34.12 &      458.0 &             0.06 &  -181.400 &         0.00022 \\
            1500 &           67.54 &              31.89 &      332.8 &             0.07 &  -181.514 &         0.00021 \\
            1500 &           72.14 &              29.86 &      236.6 &             0.09 &  -181.600 &         0.00033 \\
            1500 &           81.97 &              26.28 &      106.4 &             0.09 &  -181.705 &         0.00027 \\
            1500 &           92.65 &              23.25 &       32.6 &             0.13 &  -181.746 &         0.00034 \\
            \\
            2000 &           61.01 &              35.31 &      536.9 &             0.10 &  -181.318 &         0.00037 \\
            2000 &           63.14 &              34.12 &      461.0 &             0.08 &  -181.389 &         0.00027 \\
            2000 &           67.54 &              31.89 &      335.6 &             0.08 &  -181.504 &         0.00027 \\
            2000 &           72.14 &              29.86 &      239.2 &             0.11 &  -181.591 &         0.00034 \\
            2000 &           81.97 &              26.28 &      109.4 &             0.17 &  -181.694 &         0.00054 \\
            2000 &           92.65 &              23.25 &       35.6 &             0.22 &  -181.735 &         0.00076 \\
            \\
            2500 &           61.01 &              35.31 &      539.5 &             0.14 &  -181.308 &         0.00038 \\
            2500 &           63.14 &              34.12 &      463.7 &             0.14 &  -181.379 &         0.00051 \\
            2500 &           67.54 &              31.89 &      338.6 &             0.10 &  -181.493 &         0.00032 \\
            2500 &           72.14 &              29.86 &      242.3 &             0.17 &  -181.579 &         0.00058 \\
            2500 &           81.97 &              26.28 &      112.5 &             0.15 &  -181.683 &         0.00046 \\
            2500 &           92.65 &              23.25 &       38.7 &             0.19 &  -181.725 &         0.00053 \\
            \\
            3000 &           61.01 &              35.31 &      542.5 &             0.11 &  -181.298 &         0.00037 \\
            3000 &           63.14 &              34.12 &      466.6 &             0.20 &  -181.368 &         0.00072 \\
            3000 &           67.54 &              31.89 &      341.1 &             0.16 &  -181.484 &         0.00057 \\
            3000 &           72.14 &              29.86 &      245.6 &             0.22 &  -181.568 &         0.00075 \\
            3000 &           81.97 &              26.28 &      115.1 &             0.23 &  -181.674 &         0.00068 \\
            3000 &           92.65 &              23.25 &       41.7 &             0.45 &  -181.714 &         0.00141 \\
            \end{tabular}
        \end{ruledtabular}
\end{table*}

\subsection{Free energy model}
We fit the Helmholtz free energy ($F$) as a function of $V$ and $T$, $F(V,T)$. In
FPMD, we have direct access to the variables of volume ($V$), temperature ($T$), pressure ($P$), and 
internal energy ($U$). 
Cohen and G\"ulseren \cite{Cohen2001} studied the thermal EoS of tantalum (Ta) in full potential LAPW and mixed-basis pseudopotential methods.
 An accurate high-temperature global EoS was formed from the $T=0$ K Vinet isotherm and the thermal free-energy was
fitted by the polynomial expansion in $V$ and $T$ (see Eq. (11) in Ref.~\cite{Cohen2001}). de Koker and Stixrude \cite{deKoker2009} computed the free energy of MgO periclase and MgSiO$_3$
perovskite using FPMD, where the excess free energy was fitted in a similar expansion.
Incorporating the Debye model \cite{Moruzzi1988}, the total free energy is approximated by the polynomial expansion up to order $N_i, N_j$,
\begin{equation}
    F(V,T) = \sum_{\substack{i,j=0 }}^{N_i,N_j} A_{ij} T^{i} (V^{-\frac{2}{3}})^{j} + F_{0}\,.
    \label{eq:F_expansion}
\end{equation}
Neglecting the zero-point motion, $F_{0} = 
k_B T \left[ - D_3(x) + 3 \ln (1 - e^{- x})\right] $ where a dimensionless parameter $x = \frac{\td}{T}$ with Debye temperature $\td$.
$k_B$ is the Boltzmann constant.
$D_3(x)$ is the third order Debye function (see Appendix \ref{appendixB}). $A_{ij}$ are fitting coefficients yet to be determined. 
For comparison, we mention the classical model, where $F_{0} = - 3 k_B T \ln T$, 
with $T\ln T$ giving the proper
classical behavior at low temperatures. That is $C_V = 3 k_B$ and $S=-\infty$ at 0 K. The Debye temperature $\td$ cannot be determined from the $U(T,V), P(T,V)$
data from the classical molecular dynamics, so we obtain $\td$ from the RMS displacements (see below). For simplicity, the Debye temperature at $P=0$ GPa, $T$=300 K,
is used.

\section{Results}
\label{sec:results}
We obtained the equilibrated quantities from FPMD, where $U,P$ includes the ionic kinetic energy and ideal gas pressure, respectively. We subtracted each internal energy by the 
global minimum, as only the energy difference matters. The pressure and internal energy are $P=-\left(\frac{\partial F}{\partial V}\right)_{T}, U = F + T S = F - T\left(\frac{\partial F}{\partial T}\right)_{V}$.
$(U, P)$ data are grouped as a pair and fitted together to avoid bias between these two quantities. The fitting was performed using the weighted least-square 
fit with the {\textit{lm}} function including offset in R language. Internal energy $U$ and pressure $P$ were 
fitted simultaneously to $F(V,T)$.
$w = 1/\Delta^2$ is set for the weight, where $\Delta$ is the standard deviation of $U$ and $P$.
We fitted Eq.~(\ref{eq:F_expansion}) with $N_i = 2, N_j = 3$. 
We analyzed the MD trajectories using the code VMD, and computed the root-mean-square displacement (RMSD) for each run. From this we can obtain the effective Debye temperature $\td$ using:
\begin{equation}
    \label{eq:debye_rmsd}
    \langle u^2 \rangle = \frac{3h^2}{4\pi^2 M k_B \td} \left( \frac{D_1 (\td/T)}{\td/T} + \frac{1}{4} \right) \,,
\end{equation}
where $\vec{u}, M, h$ are  the displacement vector, the ion mass, the Planck constant, 
and $D_1$ is the first order Debye function.  The quantum correction term $\frac{1}{4}$ shall be omitted in the classical treatment.  
Since the phonon density of states is not exactly Debye-like, this is the effective Debye temperature for the second moment of the vibrational density of states (VDOS) , 
not exactly equal to the thermodynamic Debye temperature \cite{Wallace1965}.

\begin{figure}
	\includegraphics[width=1.05\linewidth]{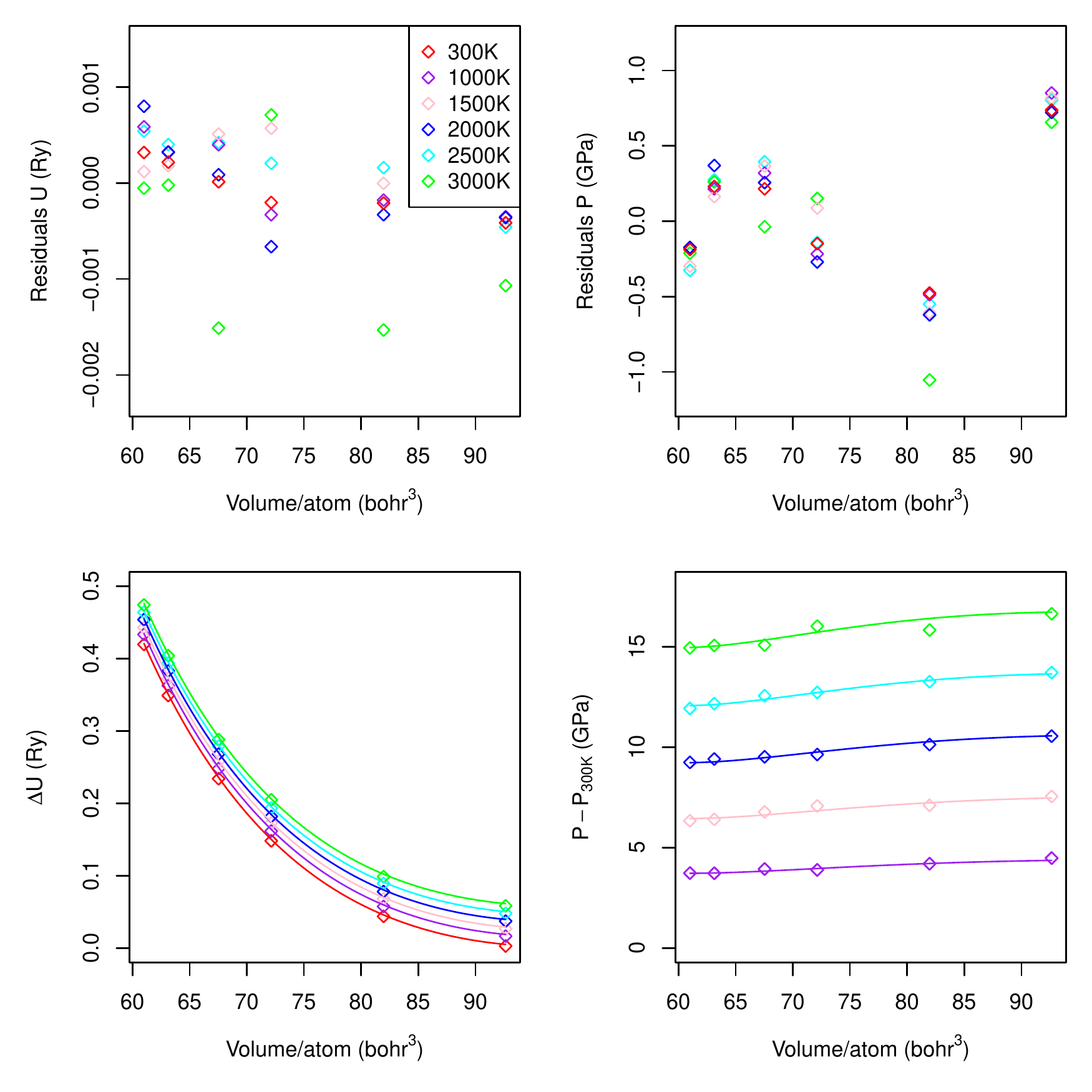}
    \caption{    \label{fig:residuals}
    The residuals of the fit to Eq.~(\ref{eq:F_expansion}) for the EoS of \fcc{} iridium are shown in a) and b). $\Delta U = U - U_{min}$, where $U_{min}$ is the minimum in 
    the dataset underlined in Table~\ref{tab:EOS_data}. The fitted curves are compared against the dataset in c) and d).
    $\Delta U$ and thermal pressure $P - P_{\mathrm{300K}}$ from the fit align well against the dataset.}
\end{figure}
The residual is the deviation between the target function and the sample mean. From Fig. \ref{fig:residuals}, we observe that the residuals are randomly 
distributed across the volume range. The absolute value of 
residuals for $U$ and $P$ are less than 0.002 Ry and 1.0 GPa (except for data point at 3000 K). For the internal energy, a global minimum $U_{min}$ is subtracted from the dataset. 
On the scale of half Ry, the internal energy is well represented. As for the pressure, we computed the pressure differences with respect to the $T=300$ K
reference and the fitted curves aligned with the dataset. Only the $T=3000$ K fit is slightly off. The resultant fitting coefficients in atomic unit for both the Debye model 
and the classical model are tabulated (see Table \ref{tab:coeffmat}).
The statistical summary from {\textit{lm}} function is included in the Appendix \ref{appendixA} (see Fig.~\ref{fig:fitSummary}).
\begin{table*}
    \caption{Coefficient matrix $A$ in atomic units for the choice of $F_0$, the Debye model and the classical model. The root mean squared errors (RMS) in the 
    fitting for the pressure $P$ (in GPa) and the internal energy $U$ (in mRy) are listed.
}
    \begin{ruledtabular}
        \label{tab:coeffmat}
    \begin{tabular}{ccccccc}
                & \multicolumn{4}{c}{$A_{ij}$}                  &      $P$ RMS (GPa)             &     $U$ RMS (mRy)     \\
                \hline
    \multirow{3}{*}{Debye model} & \multicolumn{4}{l}{\multirow{3}{*}
    {
        \begin{tabular}[c]{ccccc}
            1.904     & -62.84      & 88.81      & 8175         \\
            0         &   0.008698  & -0.1175    &    0.5574    \\
            $6.681\times10^{-09}$ &  $-3.752\times10^{-07}$ &  $6.321\times10^{-06}$ &   $-3.439\times10^{-05}$ \\
        \end{tabular}
    }} & \multirow{3}{*}{0.5380} & \multirow{3}{*}{0.706} \\
                      & \multicolumn{4}{l}{}                  &                   &                   \\
                      & \multicolumn{4}{l}{}                  &                   &                   \\
                      \hline
    \multirow{3}{*}{Classical model} 
    & \multicolumn{4}{l}{\multirow{3}{*}{
        \begin{tabular}[c]{ccccc}
            1.902     & -62.83      & 88.68      & 8176         \\
            0         &   0.008676  & -0.1171    &    0.5549    \\
            $6.567\times10^{-09}$ &  $-3.667\times10^{-07}$ &  $6.166\times10^{-06}$ &   $-3.347\times10^{-05}$ \\
        \end{tabular}  
    }} & \multirow{3}{*}{0.5378} & \multirow{3}{*}{0.661} \\
                      & \multicolumn{4}{l}{}                  &                   &                   \\
                      & \multicolumn{4}{l}{}                  &                   &                  
    \end{tabular}
        \end{ruledtabular}

\end{table*}

\subsection{$P-V-T$ EoS}
\begin{figure}
	\includegraphics[width=1.0\linewidth]{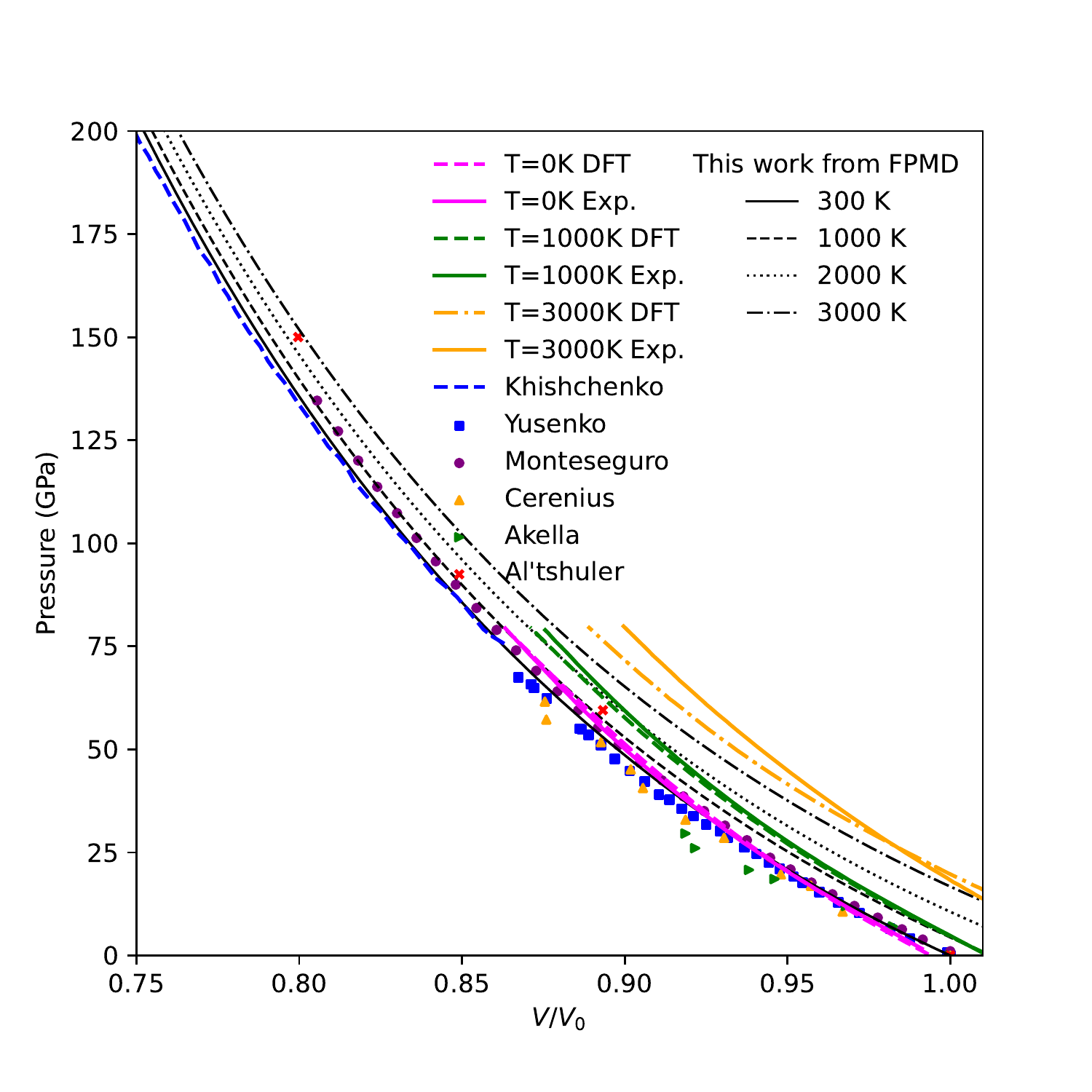}
    \caption{\label{fig:EoS} 
    Theoretical and experimental EoS's of \fcc{} iridium are compared. The DAC experimental data  
    for Yusenko \cite{Yusenko2019} Monteseguro\cite{Monteseguro2019},  Cerenius\cite{Cerenius2000}, and
    Akella\cite{Akella1982} were compared  
    at 300 K, where the BM EoS's were available. Experimental and theoretical data from Anzellini \cite{Anzellini2021} 
    are included (pink, green, and yellow lines). 
    The shockwave data (red cross) of Al’tshuler
    were taken from Ref.~\cite{Monteseguro2019}.
    Room temperature isotherm of Khishchenko shock experiments \cite{Khishchenko2022} agrees 
    very well against our FPMD results. 
    }
\end{figure}

The equilibrium atomic volume ($P=0$ GPa) at 300 K is 14.559 \AA$^3$, 2.9\% larger 
than the experimental value 14.145 \AA$^3$. The overestimation
of the lattice constants is expected for the PBE exchange-correlation functional. 
Experimental $P-V$ curves of 300 K isotherm are readily compared with our theoretical predictions.
Pressures measured  by Akella et al. \cite{Akella1982} are underestimated for compression (see Fig. \ref{fig:EoS}),  $\Delta V/V_0$   larger than 0.05 with $\Delta V=V_0 - V$.
Overall the theoretical 300 K isotherm agrees well with the experiments
 within the uncertainty especially when the compression 
is smaller than 0.15 ($P<70$ GPa) \cite{Cerenius2000,Yusenko2019,Monteseguro2019}. 
In contrast, the 3rd order BM fit done by Monteseguro et al. \cite{Monteseguro2019} sits along our 1000 K isotherm for compression $>0.15$, and reflects the inadequecy of BM EoS at high compression.   
For comparison, we have also included the FPMD and experimental study of Anzellini et al. \cite{Anzellini2021}. Their $P-V-T$ curves (both theory and experiments) 
below 80 GPa are obtained using the EoSFit7 package with ingredients such as the third-order BM EoS for the isothermal part. Their FPMD used the local
density approximations and smaller energy cutoff (300 eV).
Isotherm of 0 K compared well against our 300 K curve at low compression but not at high compression (compression $>$ 0.90). Similar 
for the isotherms of 1000 K and 3000 K.
The shock-wave experiment by Al'tshuler et al. \cite{Altshuler1969,Altshuler1981, Nemoshkalenko1988} exhibits quite distinct behavior in the
$P-V$ curve.
Around compression of 0.1,
the temperature is pinned slightly above the isotherm of 1000 K and at compression of 0.22 the temperature is close to the 3000 K isotherm. 
The high compression pressure ($\approx 600$ GPa) of Al’tshuler 
et al. was mistakenly reported in Ref.~\cite{Monteseguro2019}. 
The recent shockwave experimental work by Khishchenko \cite{Khishchenko2022} is also compared. We observe the room temperature isotherm of recent work by Khishchenko et al. align almost perfectly with 
our EoS data.
The data by Monteseguro et al. runs along our 1000 K isotherm for compression over 0.1. It is well-known that dynamic compression experiment lead to a temperature rise. Contrary to the claim that the temperature 
effect is negligible by Monteseguro et al., \cite{Monteseguro2019}
we believe the temperatures increase (not measured) along the shock
compression $P-V$ curve is significant from our predicted EoS. 

\begin{figure}
	\includegraphics[width=1.0\linewidth]{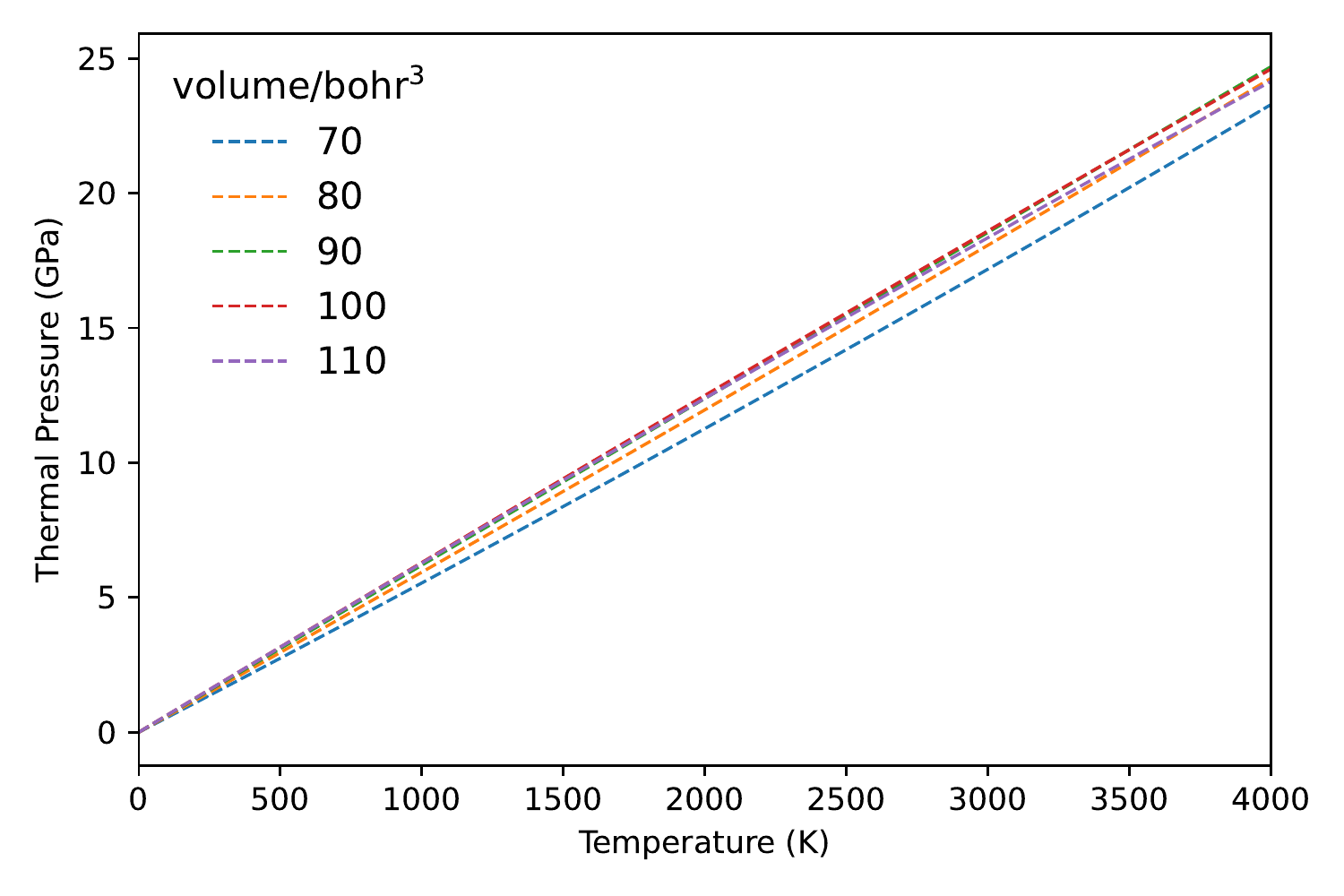}
    \caption{\label{fig:pth} The thermal pressure of \fcc{} iridium is roughly linear in $T$.}
\end{figure}
Thermal pressure measures the pressure change upon temperature increase at constant volume, $P_{th} (V, T) = P(V,T) - P(V, T_0)$.
The thermal pressure is quite linear in $T$ given that $\alpha K_T$ ($\alpha$ and $K_T$ are the thermal expansivity and the bulk modulus) is constant in the classical regime (above the Debye temperature), expressed as
\begin{equation}
    P_{th} (V, T) = \int_{T_0} ^{T}\!\! d T 
    \left( \frac {\partial P } {\partial T} \right)_V = \int_{T_0} ^{T}\!\! d T \, \alpha K_T \,.
\end{equation}
An oversimplified linear equation (see Fig.\ref{fig:pth}) could be given to the thermal pressure $P_{th}(T) = 
\lambda T$, with $\lambda = 0.0056 $ GPa/K for the equilibrium volume. One could also see the volume dependence is weak from the bottom right panel of 
Fig.\ \ref{fig:residuals}.

The equilibrium bulk modulus $B_0$ (or inverse compressibility at room temperature and zero pressure) is an important parameter in the EoS formula, such as the Vinet EoS \cite{Vinet1986,Vinet1987}.
The fitted bulk modulus is compared against earlier studies (see Table.~\ref{tab:bulk_modulus}). 
We note that Cerenius and Dubrovinsky \cite{Cerenius2000} obtained similar bulk modulus, $354$ GPa \textit{versus} $306$ GPa, by fitting the second order 
BM EoS with constraint $B_0'=4$, 
or third-order BM EoS without constraint both using experimental equilibrium volume. 
Park et al. obtained the bulk modulus of 399 GPa and 344 GPa for the LDA and GGA functional in DFT, respectively \cite{Park2015}. 
We note that $B_0$ from our fit is close to the accepted value of about 365 GPa and evidently smaller than Cynn's value 383 GPa \cite{Cynn2002}. The parameter $B_0'$ 
from our model is $5.3$.

\begin{table*}
  \caption{Experimental equilibrium volume $V_0$ (\AA$^3$ per atom), bulk modulus $B_0$ (GPa), and $B_0'$ at room temperature are compared against reported theoretical results.
  Data and method are briefly summarized, and the original references are given.
  }
      \begin{ruledtabular}
          \label{tab:bulk_modulus}
  
      \begin{tabular}{c c c c c}
          Method description   & $V_0$           & $B_0$   & $B_0'$  &  References\\
                              & (\AA$^3$/at)    & (GPa)   &    \\
          \hline
          Exp. data fitted to 3rd-order BM EoS &    14.120   &  339   & 5.3  &   Monteseguro et al. \cite{Monteseguro2019} \\
          \hline
          Exp. data fitted to 
          3rd-order BM EoS      & 14.145          &  383   & 3.1 &    Cynn et al. \cite{Cynn2002} \\
          \hline
          Exp. data fitted to &                 &       & &   Cerenius and Dubrovinsky\cite{Cerenius2000} \\
          2nd-order BM EoS, with $B_0^\prime = 4$         &  14.173 (exp. value)  &  354  & 4.0  &    \\
          3rd-order BM EoS, without constraint  &  14.173  (exp. value) &  306    &  6.8  &   \\
          \hline
          DFT data 
          fitted to BM EoS      &           &     &  &  Park et al. \cite{Park2015}, Table 1 and 2 \\
              PAW LDA      & 13.925          &  399    &  &    \\
              PAW GGA      & 14.524          &  344    &  &   \\
          \hline
          FPMD data fitted to 
              3rd-order BM EoS      & 14.150         &  366  &  5.0 &    Burakovsky et al. \cite{Burakovsky2016} \\
          \hline
          FPMD data fitted 
          to our EoS      &  14.559          &  361   &  5.3 &   This work \\
      \end{tabular}
      \end{ruledtabular}
  \end{table*}

\subsection{Shock compression} 
High pressure high temperature conditions are generated by laser heating \cite{Meng2006} or resistive heating \cite{Boehler1993} 
 in a DAC or by laser or gas gun \cite{Nellis1981} driven shock compression. Strong shocks obey the Rankine-Hugoniot,
\begin{equation}
    \label{eq:rankinehugoniot}
    U-U_0 + \frac{1}{2}(P + P_0) (V - V_0) = 0 \,.
\end{equation}
Since the analytical expression for $U, P$ as a function of $V, T$ is known, for each volume $V$, we solve
Eq.~(\ref{eq:rankinehugoniot}) by searching its root $T$ given the experimental value $V_0, T_0$. 

\begin{figure}
	\includegraphics[width=1.00\linewidth]{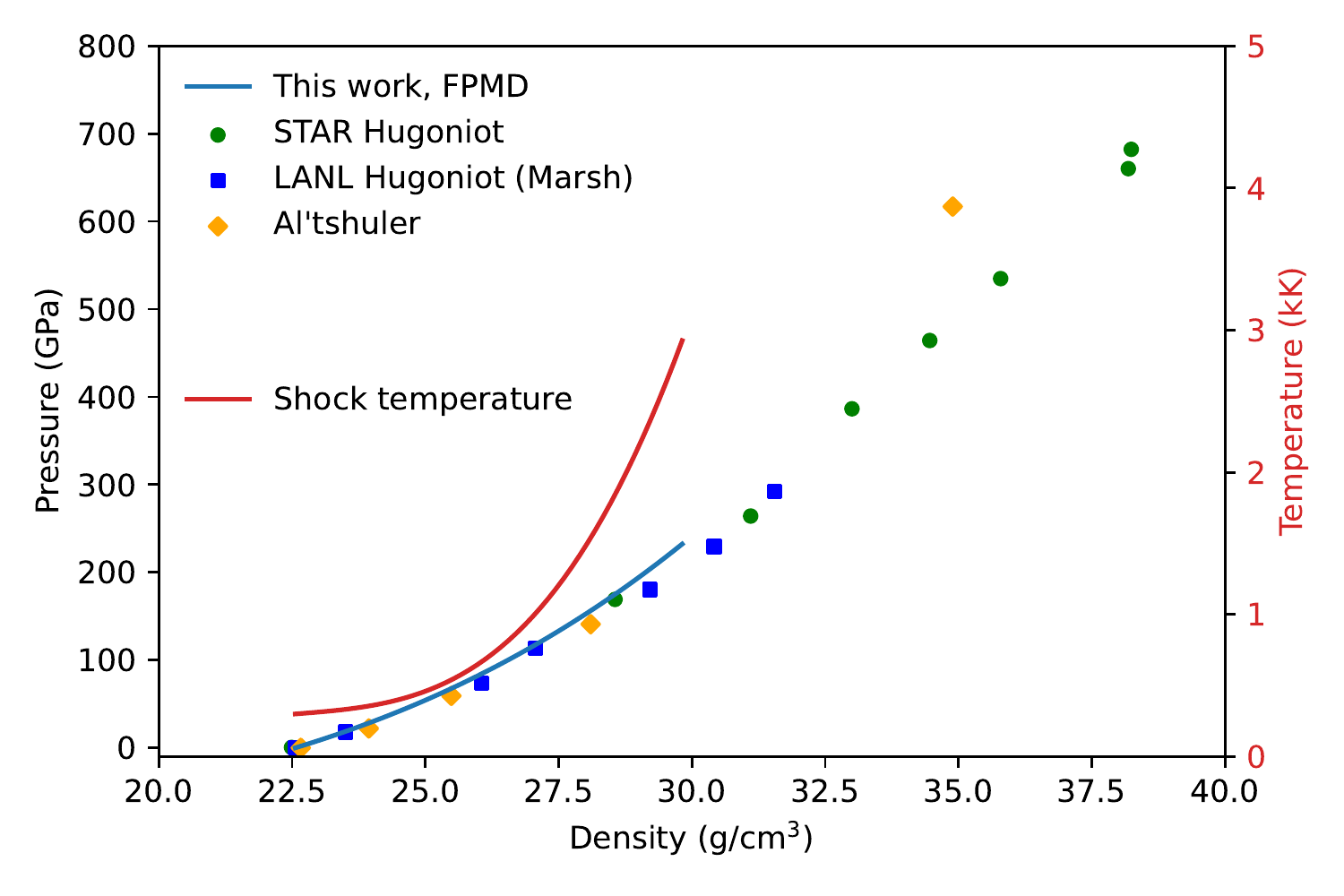}
    \caption{ \label{fig:hugoniot}
    The principle Hugoniot curve from our theoretical EoS (solid black) of \fcc{} Ir is compared against the shockwave experimental data (symbols).
    Round green is for the STAR Hugoniot \cite{Seagle2019}, square blue for the LANL Hugoniot \cite{LASLbook}, and diamond orange for Al'tshuler 
    \cite{Altshuler1969, Altshuler1981}, respectively. Computed temperature from the Eq.~(\ref{eq:rankinehugoniot}) is shown in red. The dashed extension is beyond the simulation 
    domain. 
    }
\end{figure}

We compared our predicted principle Hugoniot with available shock experimental data from several facilities
 \cite{Fortov2004,Seagle2019} (Fig. \ref{fig:hugoniot}). 
Our theoretical principle Hugoniot agrees well with that from earlier data of Al'tshuler and LANL March, 
as well as more recent data of STAR Hugoniot, for $P< 200$ GPa. Above 200 GPa, our predicted pressure is higher than
that of LANL and STAR but lower than Al'tshuler's. Our theoretical Hugoniot below 3000 K (shock temperature) are fairly reliable
which correspond to pressure less than 200 GPa. 
The shock temperature, calculated as the solution to Eq.~(\ref{eq:rankinehugoniot}), is shown.

\subsection{Equation of state parameters}
\begin{figure}
	\includegraphics[width=1.00\linewidth]{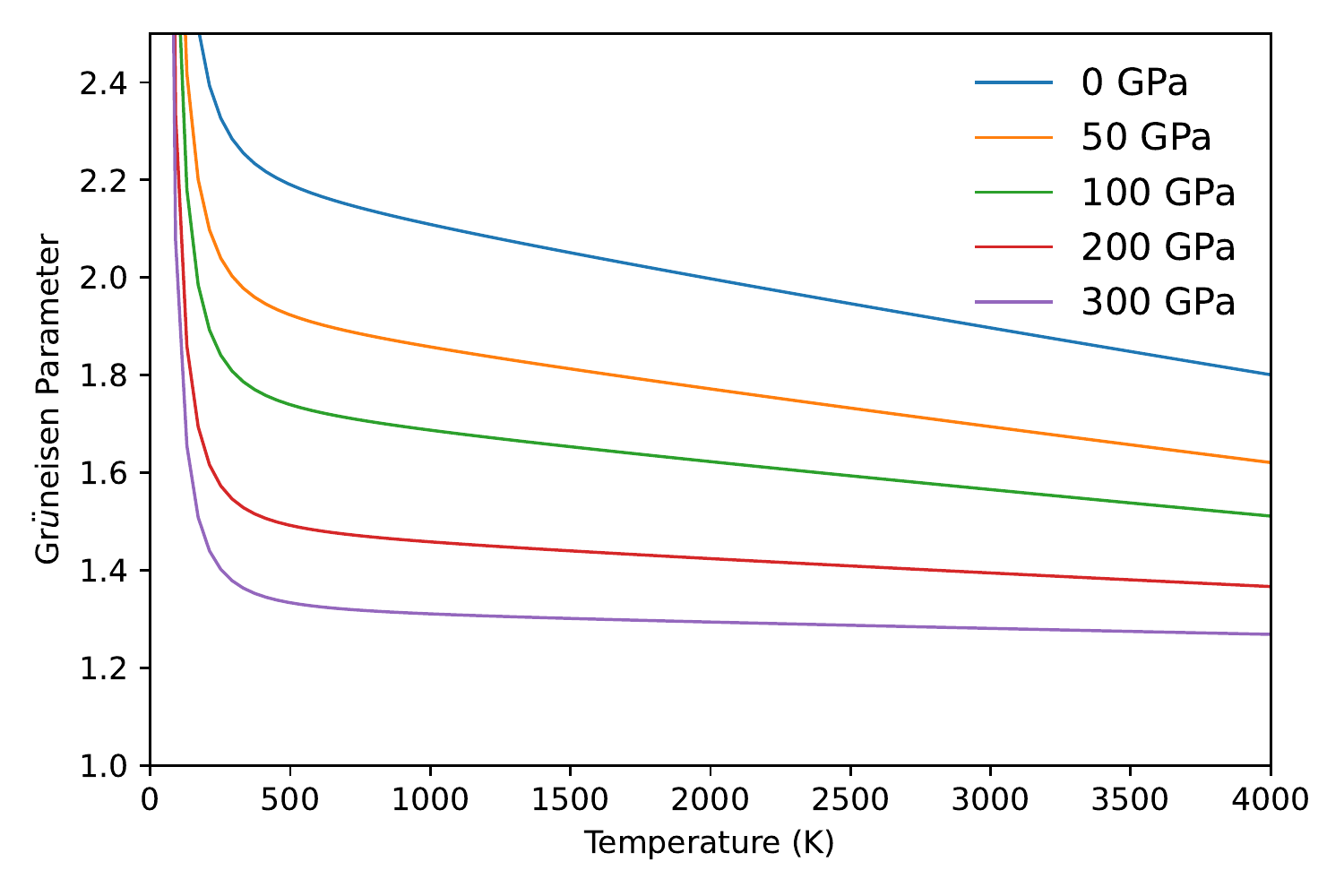}
    \caption{   \label{fig:Gruneisen}
    Gr\"uneisen parameter  $\gamma$ as a function of temperature for various pressures.
    }
\end{figure}

Thermal EoS parameters such as thermal expansivity $\alpha$, isothermal compressibility $\beta_T$, Gr\"uneisen parameter $\gamma$, and the heat capacity $C_V$ and $C_P$ are obtained
by differentiation and algebraic manipulation of Eq.~(\ref{eq:F_expansion}). We now discuss some of these parameters.
\begin{figure}
	\includegraphics[width=1.00\linewidth]{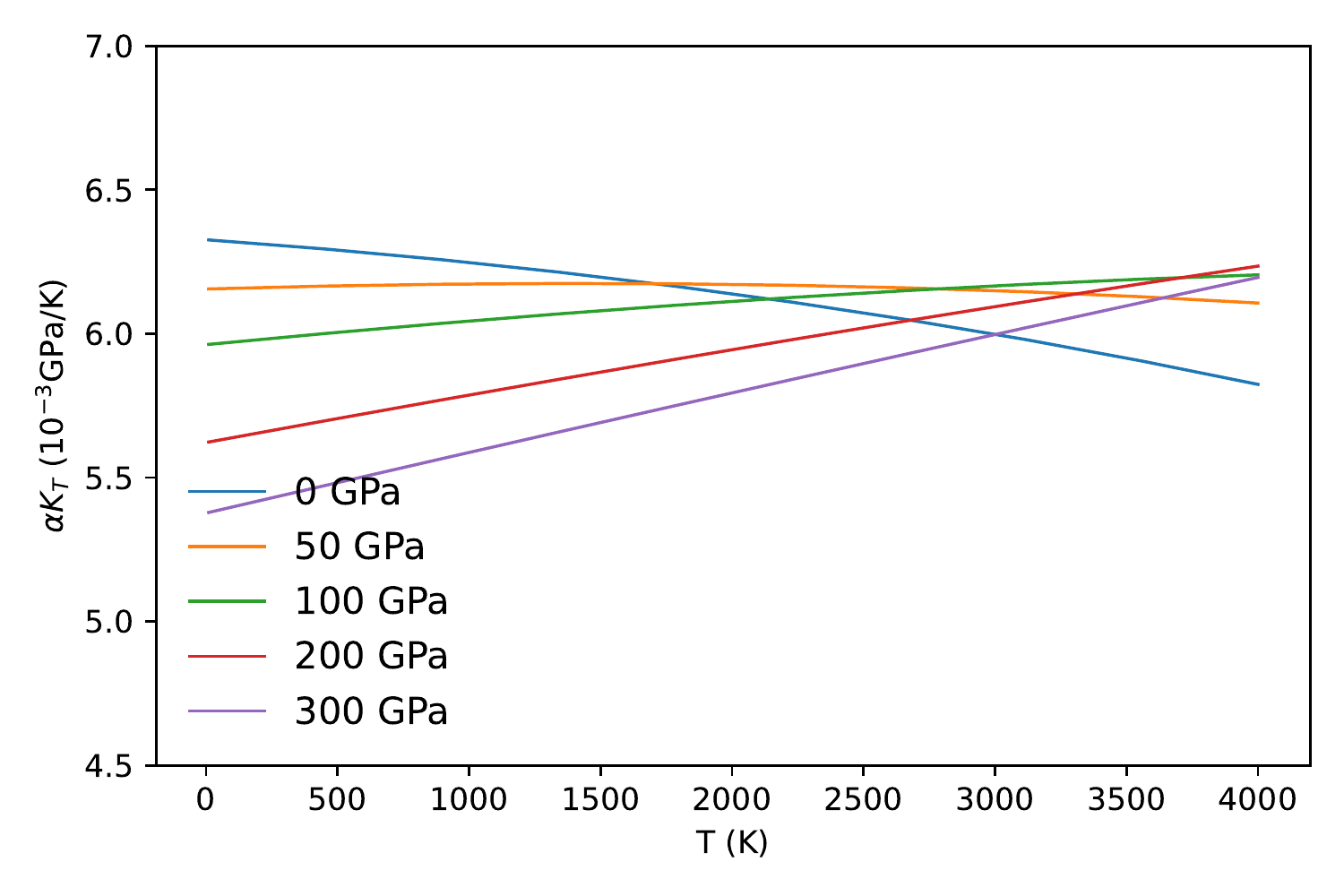}
    \caption{   \label{fig:alphaKT}
    Parameter $\alpha K_T$ as a function of temperature for various pressures.
    }
\end{figure}
As expected from the thermal pressure, $\alpha K_T$ is weakly dependent on the volume and temperature (see Fig. \ref{fig:alphaKT}). The Gr\"uneisen parameter
\begin{equation}
    \gamma = V \left (\frac{\partial P} {\partial U} \right)_V = V\frac{\alpha K_T}{C_V}
\end{equation}
is another important parameter. It is used in the Mie-Gr\"uneisen EoS, where $\gamma$ is assumed independent of temperature. The span of $\gamma$ as a function of temperature reduces when the pressure 
increases (see Fig. \ref{fig:Gruneisen}). For high compressions, it is indeed fairly temperature independent.

\begin{figure}
	\includegraphics[width=1.00\linewidth]{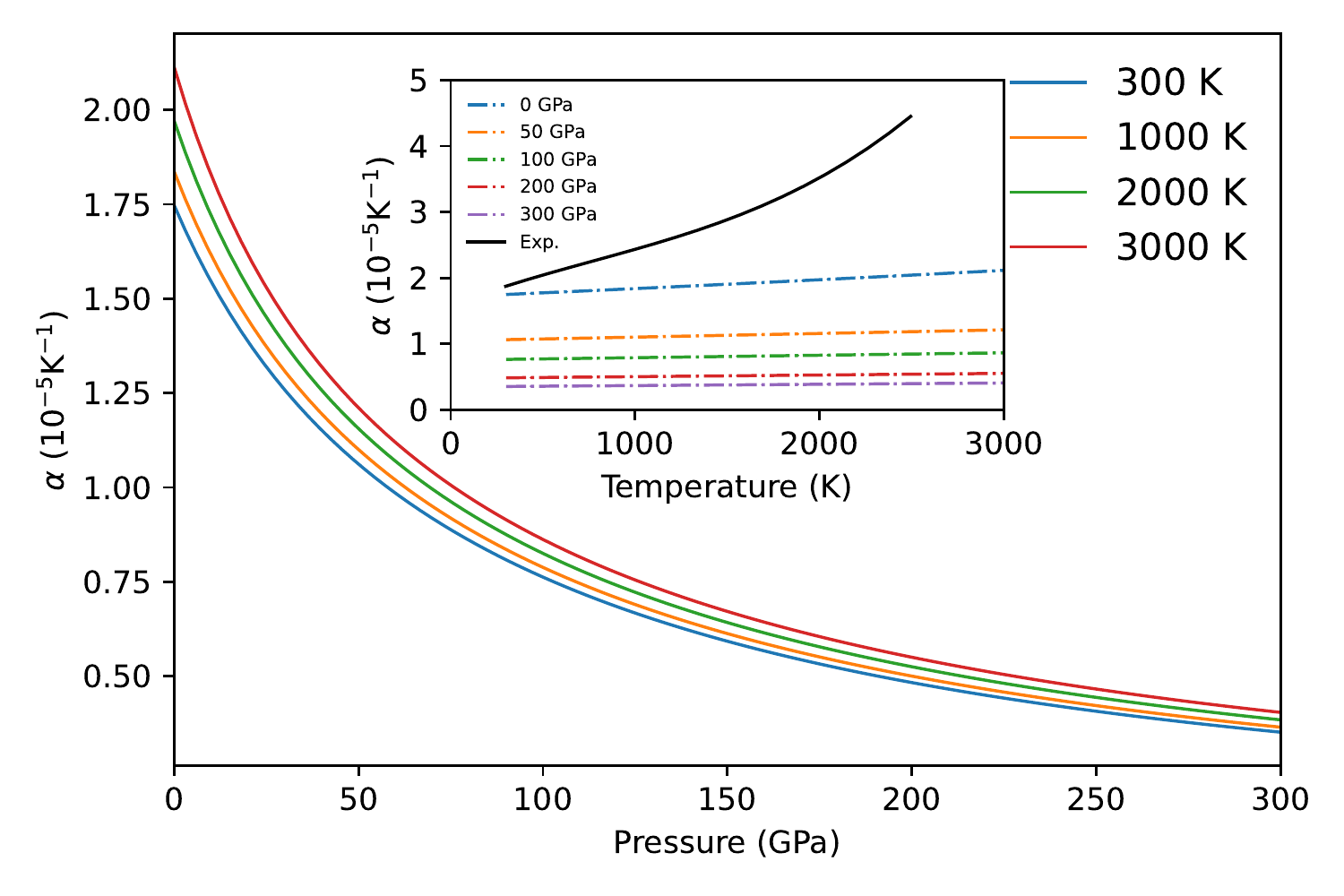}
    \caption{ \label{fig:alpha} The volumetric thermal expansivity $\alpha$ of \fcc{} iridium as a function of pressure at various temperatures. 
    Halvorson and Wimber measured the linear thermal expansion as
    $\alpha_t = a_0 + a_1 t + a_2 t^2 + a_3 t^3$ with $a_0 = 6.167 \times 10^{-6}, a_1 = 3.038 \times 10^{-9},a_2 = -0.8448 \times 10^{-12},a_3 = 0.5852\times 10^{-15}$,
    for $t$ expressed in $^{\circ}$C \cite{Halvorson1972} at ambient pressure (see inset solid line), where one can show $\alpha = 3 (L_0/L_t) \alpha_t$ for isotropic materials
    with reference length $L_0$.
    }
\end{figure}
The volumetric thermal expansivity $\alpha = - 1/V (\partial V /\partial T)_P$ for isotropic materials is three times the linear thermal expansivity
coefficient $\alpha_L$, $\alpha = 3 \alpha_L$. $\alpha$ in Fig.~\ref{fig:alpha} 
is essentially temperature independent but rather volume sensitive. Our theoretical prediction is below the 
reported experimental value \cite{Halvorson1972}, but it is noted that around room temperature, our theory prediction 
gives the right thermal expansion coefficient. The expansivity has downsized by a factor of 4 when the pressure goes to 300 GPa.
\begin{figure}
	\includegraphics[width=1.00\linewidth]{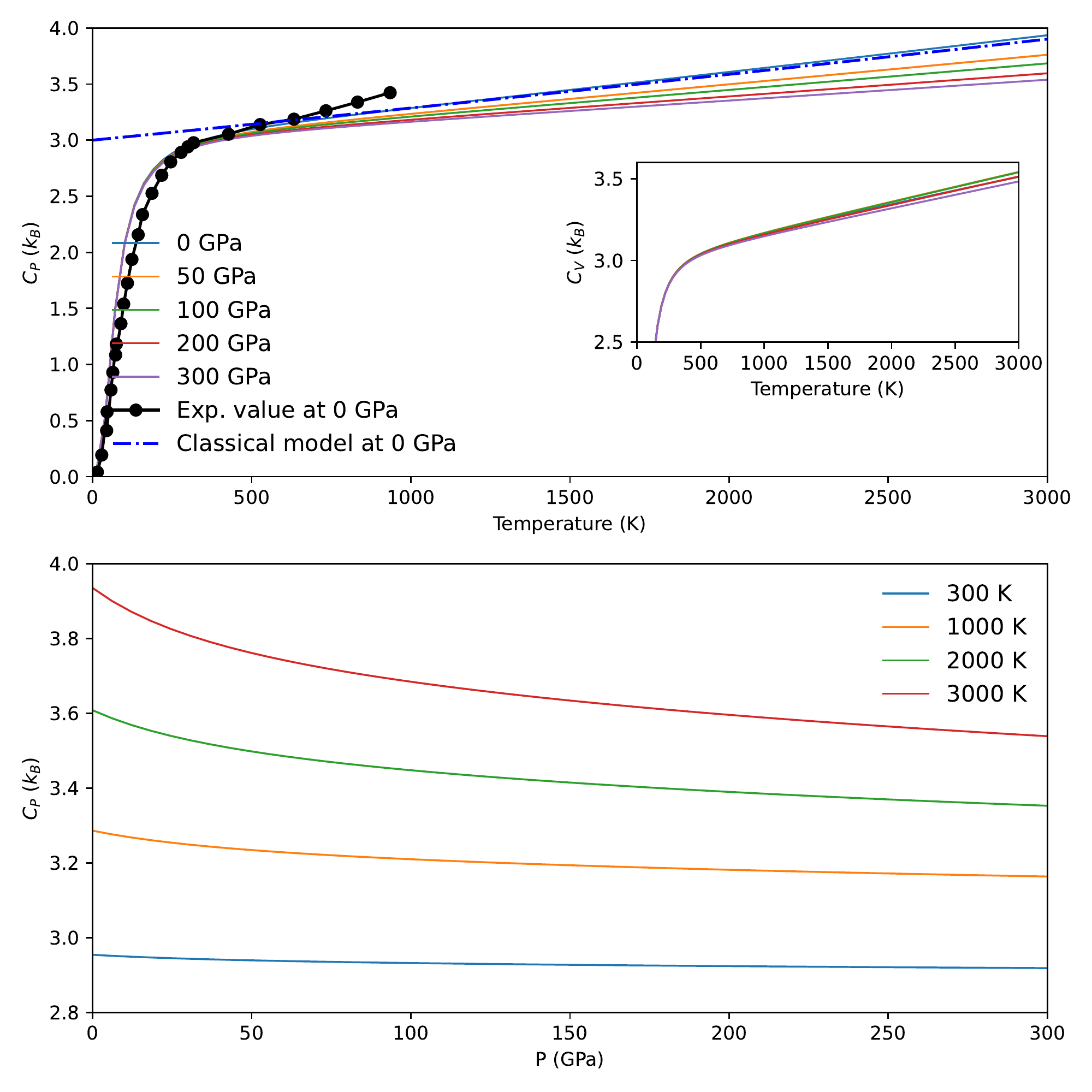}
    \caption{  \label{fig:healcapacity}
    Constant pressure heat capacity $C_P$ as a function of temperature (top panel) and pressure (bottom panel). Experimental value at ambient condition is 25.10 \jkmol \cite{Lide2004crc}. 
    The classical model (blue dash-dotted) for 0 GPa starts to deviate for $T$ below 500 K, and approaches the classical limit as $T\to 0$ K. Inset shows the constant volume heat capacity $C_V$. Heat capacity largely reduces when pressure goes up.}
\end{figure}
The heat capacity at high pressures are almost linear above 500 K, see Fig.~\ref{fig:healcapacity}. The higher the temperature, 
the slope of $C_P$ is smaller. At ~0 GPa, the predicted value 25.68 \jkmol  is fairly accurate and only 
2.3\% larger than the experimental heat capacity  25.10 \jkmol \cite{Lide2004crc}, given that we used the 
formula, $C_P = C_V (1+T\alpha \gamma)$ (see Appendix \ref{appendixA}) with errors in $C_V, \alpha, $ and $\gamma$.

The RMSD is a critical quantity for the analysis of the phonon vibrations. Moseley et al. \cite{Moseley2020} presented
 temperature-dependent inelastic neutron scattering
(INS) experiments as well as quasi-harmonic density functional theory calculations to study the thermodynamic properties of Ir. 
Our FPMD  RMSD at 300 K agrees particularly
well with their experimental findings. Their reported  $\langle u^2\rangle$ at higher temperatures (T=673 K and 823 K, see Table I of Ref.~\cite{Moseley2020})
however, are higher than our FPMD predictions. This is reasonable since our NVT ensembles at these temperatures lead to higher pressure and confined vibrations. 
It is worth to note that their phonon density of states (PDOS) integrates to 1, and is not fitted well at higher energies. Further investigation of PDOS with a
FPMD simulation to compare against the experiments may give insight to the anharmonic effects.
Debye temperatures of isochores using Eq.~(\ref{eq:debye_rmsd}) exhibit weak temperature dependence(see Fig. \ref{fig:Debye}).
\begin{figure}
	\includegraphics[width=1.00\linewidth]{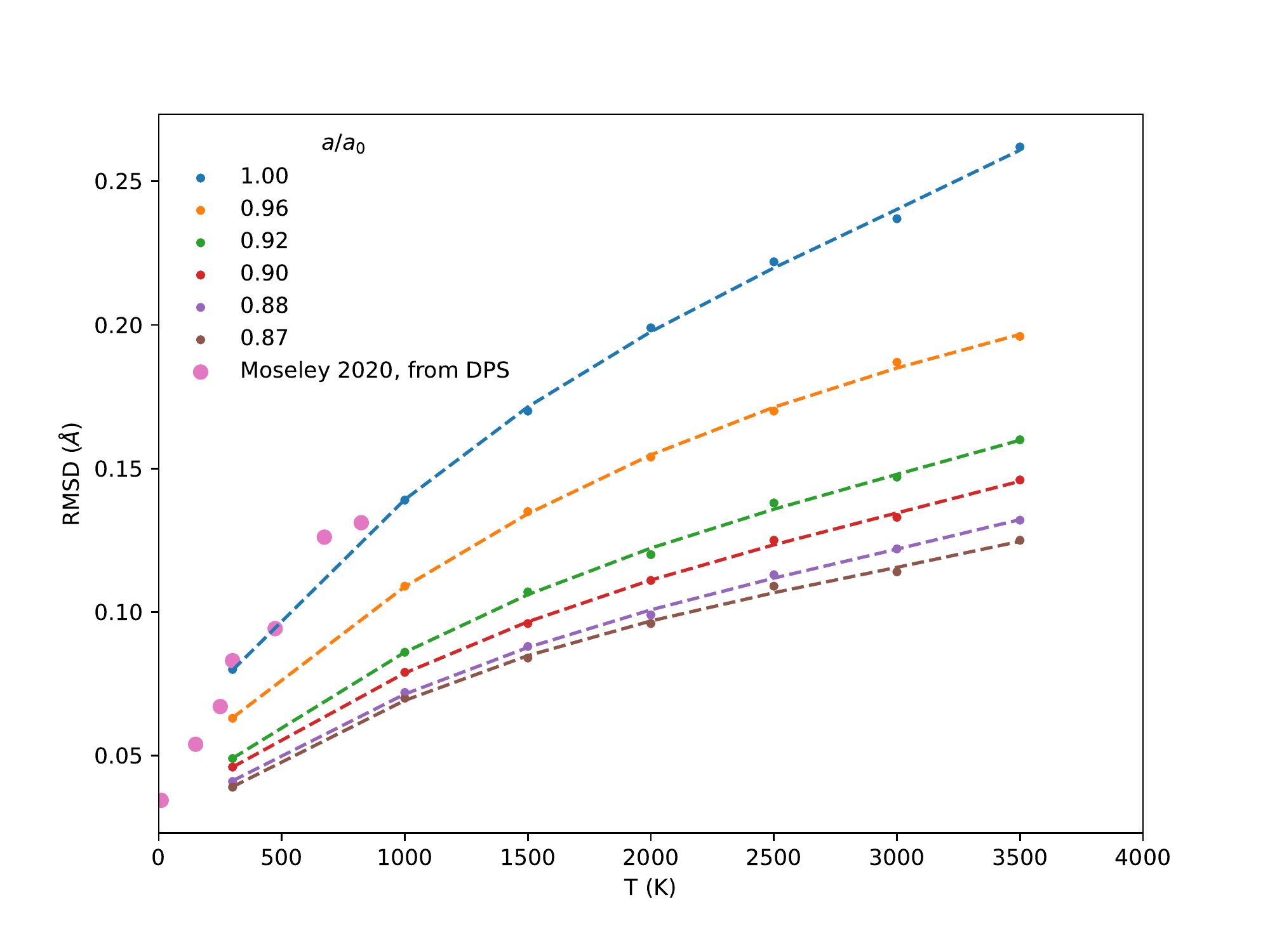}
    \caption{  \label{fig:RMSD}
    RMSD  as a function of temperature. Experimental data was obtained by Moseley et al. \cite{Moseley2020} for constant pressure.
    }
\end{figure}

\begin{figure}
	\includegraphics[width=1.00\linewidth]{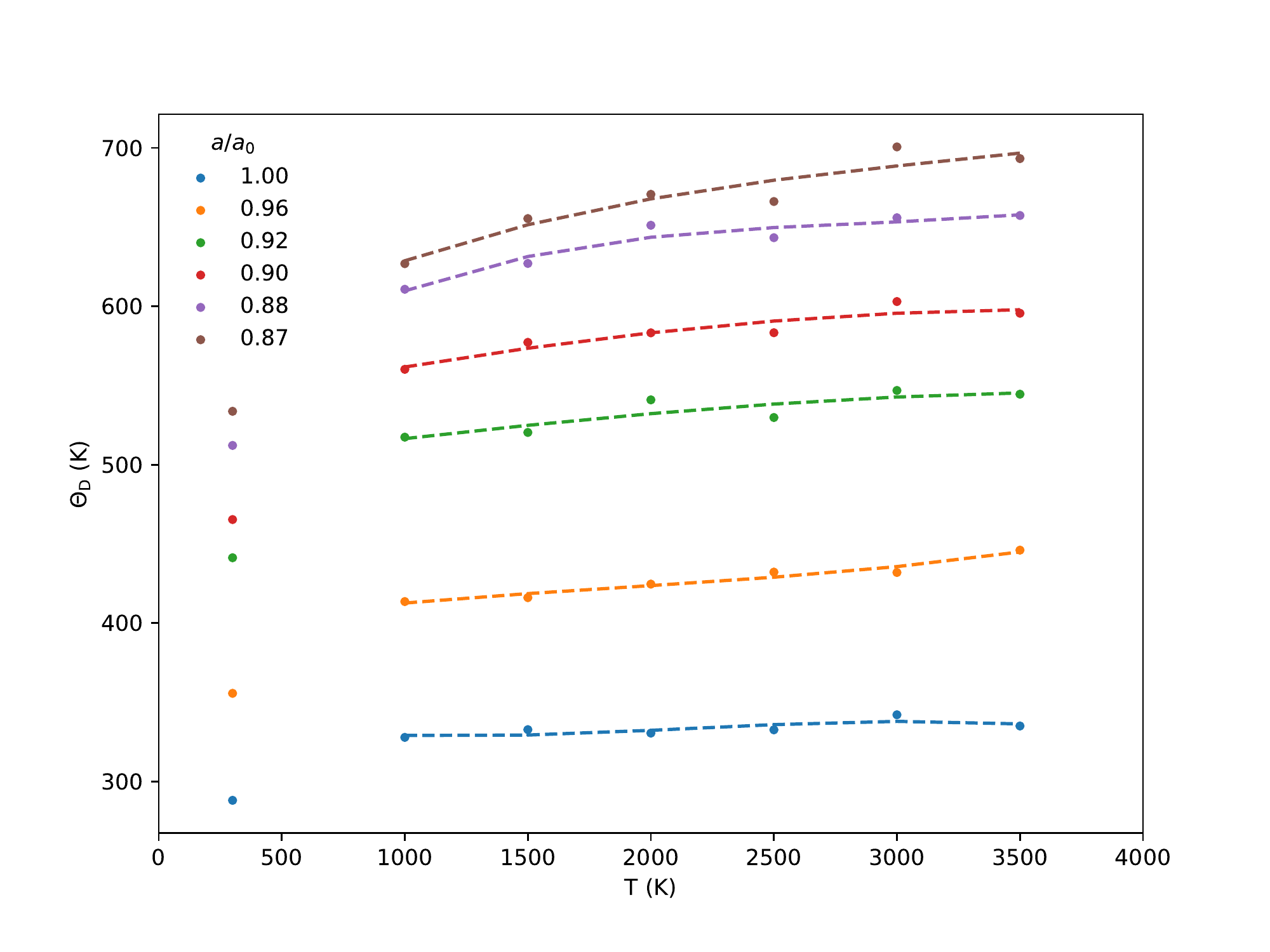}
    \caption{  \label{fig:Debye}
    Debye temperature  $\td$ as a function of temperature. 
    Zero point motion is not included to obtain the Debye temperature.     
    In the high temperature, classical region, our RMSD are classical from classical FPMD (Fig. \ref{fig:RMSD}) 
    but when an effective classical Debye temperature is derived to model the RMSD, 
    there is a large change with decreasing temperature into to quantum regime in the Debye model. 
    A 4th order polynomial fit is done only for higher temperature due to the classical treatment to the ions.
    }
\end{figure}
    
\section{Conclusions}
We have performed a series of FPMD simulations for the \fcc{} Ir at conditions up to 3000 K and 540 GPa. 
By using a simplified model for the free-energy as a function of temperature and volume and the statistical average quantities 
internal energy and pressure $(U, P)$, the thermal EoS is obtained by globally fitting to the model. We have compared previous experimental EoS's and provided 
the thermal EoS up to 3000 K, and 540 GPa. The $P-V-T$ curve is reasonably agreeing with the fitted BM EoS at low compression but differs at high compression.
Our first-principles EoS accords with the most recent shockwave experiment work by Khishchenko 
 We find that $\alpha K_T$ and the thermal pressure is quite constant from its dependence in temperature and 
which turns out to be true for a wide range of materials. We have shown some representative derived thermal parameters against available experiments and found agreements
and discrepancies. Further work might resolve these discrepancies.


\section{Acknowledgments}
The work was done under the auspices of the US National Science Foundation CSEDI grant EAR-1901813 to R.E.C. and 
the National Natural Science Foundation of China (Grant No. 12104230) to K.L.;
R.E.C. is supported by the Carnegie Institution for Science and gratefully acknowledges 
the Gauss Centre for Supercomputing e.V. for funding this project by providing computing 
time on the GCS Supercomputer Supermuc-NG at Leibniz Supercomputing Centre.
All the FPMD calculations were performed on Supermuc-NG.

\appendix 

\section{Thermodynamic relations}
\label{appendixA}
Once the Helmholtz free energy $F = F(V, T)$
is known as a function of volume ($V$) and temperature ($T$), the following
thermodynamical quantities can be obtained from it \cite{Callen1998Book}:
\begin{eqnarray}
    P&=&-\left(\frac{\partial F}{\partial V}\right)_{T} \,,\\
    S&=&-\left(\frac{\partial F}{\partial T}\right)_{V} \,,\\
    K_{T}&=&\beta_{T}^{-1}=-V\left(\frac{\partial P}{\partial V}\right)_{T}=V\left(\frac{\partial^2 F}{\partial V^2}\right)_{T} \,,\\
    C_{V}&=&T\left(\frac{\partial S}{\partial T}\right)_{V}=-T\left(\frac{\partial^{2} F}{\partial T^{2}}\right)_{V} \,,\\
    \alpha K_{T}&=&-\left(\frac{\partial^{2} F}{\partial T \partial V}\right) \,,\\
    \gamma&=&V\left(\frac{\partial P}{\partial U}\right)_{V}=V \frac{\alpha K_{T}}{C_{V}} \,,\\
    \frac{C_{P}}{C_{V}}&=&\frac{K_{S}}{K_{T}}=1+T \alpha \gamma \,,\\
    U&=&F+T S \,,
\end{eqnarray}
where pressure, entropy, isothermal compressibility (its inverse is the bulk modulus $K_T$), constant volume molar heat capacity, volumetric expansion coefficient, Gr\"uneisen parameter, internal energy
are denoted by $P, S, \beta_T, C_V, \alpha, \gamma, U$. $C_P$ is the  constant pressure molar heat capacity.
The parameter $B_0'$ (or $K_{0}' = \left.\frac{\partial K}{\partial P}\right|_{P=0}$) can thus be computed using above relations:
\begin{eqnarray}
    K' &=& \frac{\partial K}{\partial P}  \nonumber
    \\
    &=& \left(\frac{\partial K}{\partial T} \right)_V \left(\frac{\partial T}{\partial P} \right)_V + \left(\frac{\partial K}{\partial V} \right)_T \left(\frac{\partial V}{\partial P} \right)_T \nonumber 
    \\
    &=& \left(\frac{\partial K}{\partial T} \right)_V \frac{1}{\left(\frac{\partial P}{\partial T} \right)_V} + \left(\frac{\partial K}{\partial V} \right)_T \frac{1}{\left(\frac{\partial P}{\partial V} \right)_T} \nonumber 
    \\
    &=& \left(\frac{\partial K}{\partial T} \right)_V \frac{1}{\left(\frac{\partial P}{\partial T} \right)_V} - \left(\frac{\partial K}{\partial V} \right)_T \frac{V}{K_T} \,.
\end{eqnarray}

\begin{figure}
	\includegraphics[width=1.00\linewidth]{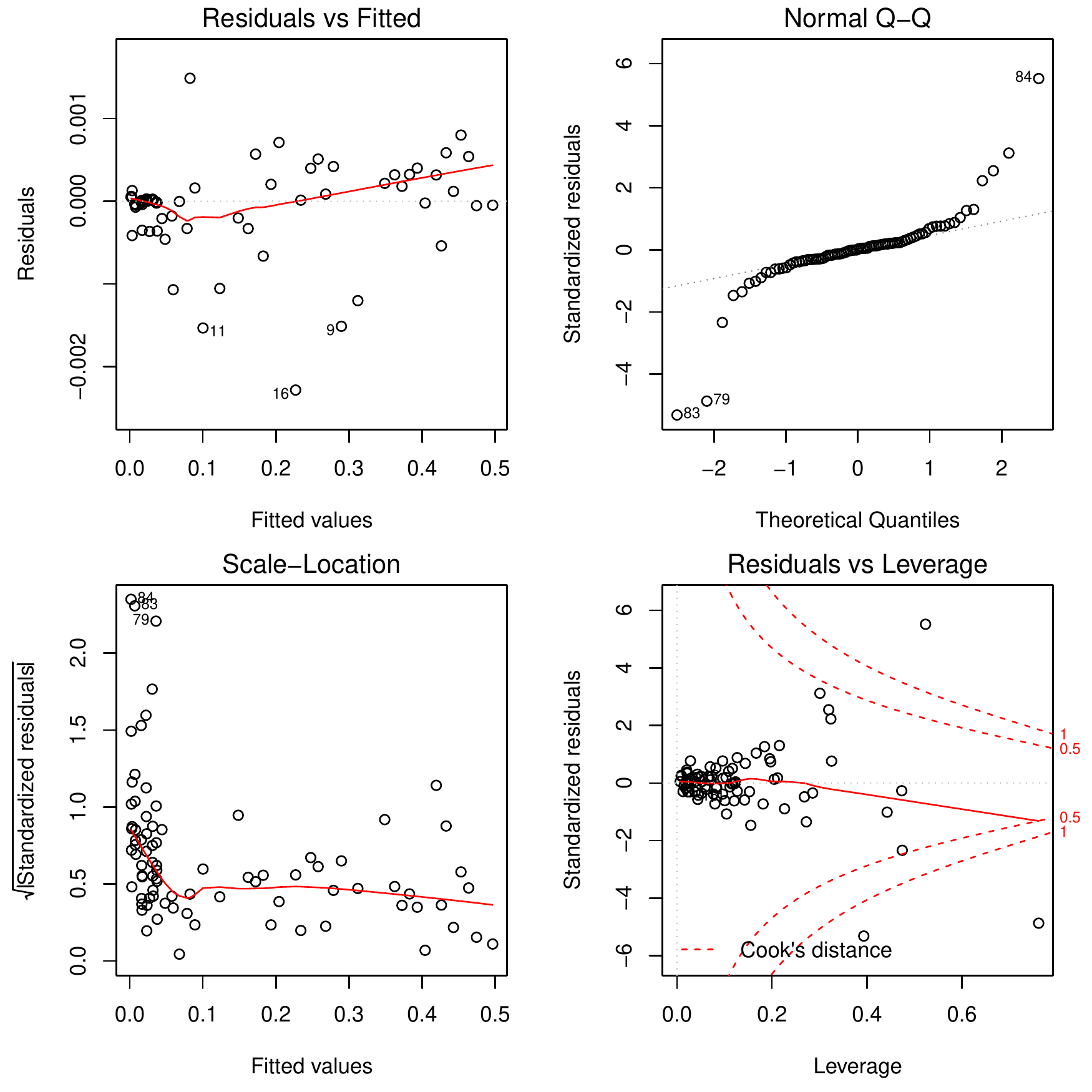}
    \caption{ \label{fig:fitSummary}
    The fit summary plot from the \textit{lm} function in R.
    }
\end{figure}



\section{Debye Model}
\label{appendixB}
The Helmholtz free energy $F$ of a vibrating lattice at volume $V$ and temperature $T$, can be approximated as
\begin{equation}
    F(V,T) = E(V) + F_{\vib}(V,T) + F_{el}(V,T)\,,
\end{equation}
where $F_{\vib}$ is the vibrating energy of the lattice and $F_{el}$ is the thermal electronic free energy which is typically negligible. Moruzzi et al. \cite{Moruzzi1988} proposed 
an empirical Debye model with 
\begin{equation}
    F_{\vib} =  k_B T \left[ - D_3(x) + 3 \ln (1 - e^{- x})\right] + \frac{9}{8}k_B \td \,,
\end{equation}
with Debye temperature $\td$ and dimensionless parameter $x = \frac{\td}{T}$.  The last term is the zero-point energy.  $D_3(x)$ is the third order Debye function.
The $n$th order Debye function is defined as
\begin{equation}
    D_n(x) = \int_{0}^{x} \frac{t^n}{e^{t} - 1} dt, x \geq 0\,,
\end{equation}
where $n$, a non-negative integer, is the order of the Debye function. 
The vibrational entropy is
\begin{equation}
    S_{\vib} = 4k_B D(x) - 3 k_B \ln (1 - e^{- x}) \,.
\end{equation}
Neglecting the zero-point motion, 
the vibrational internal energy  $U_{\vib}$ thus can be 
obtained 
\begin{equation}
    U_{\vib} = F_{\vib} + T S_{\vib} = 3 k_B  T D(x) \,.
\end{equation}

\bibliographystyle{apsrev4-1}
\bibliography{ref}

\end{document}